\documentclass[useAMS,usenatbib]{mn2e}

\usepackage{amsmath}
\usepackage{comment}
\usepackage{graphicx}
\usepackage{rotating}
\usepackage{color}
\usepackage{aas_macros}
\usepackage{amsfonts}
\usepackage{amssymb}
\newcommand{\beq}{\begin{equation}}
\newcommand{\enq}{\end{equation}}
\newcommand{\beqa}{\begin{eqnarray}}
\newcommand{\enqa}{\end{eqnarray}}
\newcommand{\beit}{\begin{itemize}}
\newcommand{\enit}{\end{itemize}}
\newcommand{\bem}{\begin{pmatrix}}
\newcommand{\enm}{\end{pmatrix}}

\newcommand{\lp}{\left (}
\newcommand{\rp}{\right )}

\newcommand{\bes}{\begin{sideways}}
\newcommand{\ees}{\end{sideways}}

\renewcommand{\bbeta}{\boldsymbol {\eta}}
\renewcommand{\bmu}{\boldsymbol {\mu}}

\title[Improving constraints on the neutrino mass using sufficient
statistics]{Improving constraints on the neutrino mass using sufficient statistics}
\author[Wolk et al.]{M. Wolk$^1$,\thanks{E-mail:
wolk@ifa.hawaii.edu}, I. Szapudi$^1$, J. Bel$^2$, C. Carbone$^{2,3}$, and J. Carron$^{1,4}$\\
$^1$Institute for Astronomy, University of Hawaii, 2680 Woodlawn
Drive, Honolulu, HI, 96822\\
$^2$INAF-Osservatorio astronomico di Brera, via Emilio Bianchi 46, I-23807 Merate (LC), Italy\\
$^3$INFN -Istituto Nazionale di Fisica Nucleare, Sezione di Bologna,
viale Berti Pichat 6/2, I-40127 Bologna, Italy \\
$^{4}$Department of Physics and Astronomy, University of Sussex, Brighton BN1 9QH, U.K}
\begin{document}

\date{\today}

\pagerange{\pageref{firstpage}--\pageref{lastpage}} \pubyear{2015}

\maketitle

\label{firstpage}

\begin{abstract}
We use the ``Dark Energy and Massive Neutrino Universe'' (DEMNUni) simulations to compare the constraining power of ``sufficient statistics'' with the standard matter power spectrum on the sum of neutrino masses,  $M_\nu \equiv \sum m_\nu$.  In general, the power spectrum, even supplemented with higher moments of the distribution, captures only a fraction of the available cosmological information due to correlations between the Fourier modes.  In contrast, the non-linear transform of sufficient statistics, approximated by a logarithmic mapping $A=\ln(1+\delta)$, was designed to capture all the available cosmological information contained in the matter clustering; in this sense it is an optimal observable.
Our analysis takes advantage of the recent analytical model developed
by \cite{Carronetal14d} to estimate both the matter power
spectrum and the $A$-power spectrum covariance matrices.
Using a Fisher information approach, we find that using sufficient statistics increases up to
$8$ times the available information on the total  neutrino mass at $z=0$, thus
tightening the constraints by almost a factor of $3$ compared to the
matter power spectrum.
\end{abstract}


\section{Introduction}
The interest for neutrino science has been stimulated over the last decade by
solar, atmospheric and accelerator experiments that showed, through
the observations of the so-called ``neutrino flavour oscillations'' that
neutrinos have a finite mass \citep{Ahmedetal04, Eguchietal03,
  Arakietal05, McKeownetal04}. These experiments are, however, only
sensitive to mass square differences between neutrino mass eigenstates,
leaving the knowledge of the total neutrino mass as one of the great unsolved
problems of modern physics. Cosmological probes, such as
the cosmic microwave background (CMB) or large scale structures (LSS) in the Universe
have the highest experimental sensitivity to date to the absolute neutrino mass, as massive neutrinos influence
structure formation \citep{Dolgov02, Lesgourguesetal06, Lesgourguesetal12}.

The epoch at which massive neutrinos became non-relativistic changes
the time of matter-radiation equality, increasing slightly the size of
the sound horizon at recombination and thus changing the position
of the CMB anisotropy peaks. In combination with Baryon Acoustic Oscillations (BAO), 
measurements from the Planck satellite have provided the tightest
constraints on the total neutrino mass $M_\nu< 0.17$ eV at
$95\%$ CL \citep{Planck15}.  However this result is sensitive to the value of the
Hubble parameter and including other external measurements leads to the
weaker, more conservative limit of $M_\nu < 0.23$ eV.
Neutrinos also impact the growth of structure by
suppressing the small-scale matter density fluctuations. Thus 
measurements of the total matter power spectrum can improve constraints on
the neutrino mass in a complementary way.
The Sloan Digital Sky Survey (SDSS) DR8
LRG angular power spectrum with the WMAP7 data and a HST prior on the
Hubble parameter gives $M_\nu < 0.26$ eV ($95\%$ CL)
\citep{dePutteretal12}. More recently \cite{Beutleretal14}, using
the Baryon Oscillation Spectroscopic Survey (BOSS) CMASS Data Release
11 combined with measurements from the Planck satellite (without the
$A_{L}$-lensing signal), weak lensing data and BAO constraints, found
$M_\nu = 0.36 \pm 0.10$ eV ($68\%$ CL).

These bounds mainly use structure formation data up to the translinear regime
($k_{max}\sim 0.15$ $h$Mpc$^{-1}$), but even on such large scales some non-linear contamination is present
and must be taken into account.
Moreover, future galaxy redshift surveys such as the Dark Energy
Survey\footnote{http://www.darkenergysurvey.org/} (DES), the Large Synoptic
Survey Telescope\footnote{http://www.lsst.org/lsst/} (LSST),
Euclid\footnote{http://sci.esa.int/euclid/} or the Wide Field
Infra-Red Survey Telescope\footnote{http://wfirst.gsfc.nasa.gov/}
(WFIRST), are expected to constrain the matter power
spectrum at percent level precision at least on scales of $k \sim 0.1-1$
$h$Mpc$^{-1}$ \citep{Hearinetal12}.

Yet, the power spectrum, as well as higher order statistics, lose their
effectiveness beyond linear scales due to the emergence of coupling among
Fourier modes induced by the non-linear gravitational evolution
\citep{Rimesetal05, Neyrincketal06}. It means that even if these future
surveys can probe more deeply the non-linear regime, the amount of
accessible information on the neutrino mass, or in general on other cosmological parameters, will be drastically reduced.
Non-linear transformations, such as the logarithmic
mapping \citep{Neyrincketal09} or variants thereof
\citep{Seoetal11,Joachimietal11} were introduced specifically to
recapture this hidden
information. \cite{Carronetal13}, using
perturbation theory, shows that this logarithmic
transformation, $A = \ln(1 + \delta)$, approximates well the exact
``sufficient statistics''  for a continuous field, meaning that it captures, by
design, all the available cosmological information from the matter field.
With the data analysis of future cosmological surveys in mind, investigating
the constraining power of this new observable via $N$-body simulations with a massive neutrino component  is necessary. Our goal is to extract all possible information on cosmological parameters, especially, the neutrino mass, the focus of the present work.

We proceed as follows. Section~\ref{sec:simu} presents the DEMNUni
simulations. The analytical model used for the estimation of the matter power
spectrum and the $A$-power spectrum covariance matrices is described
in Section~\ref{sec:results}.
Making use of this model, we forecast the Fisher information on the
total neutrino mass for both the non-linear transform $A$ and the
matter power spectra. We
summarize and conclude with a discussion in Section~\ref{sec:discuss}.
\section{Simulations}
\label{sec:simu}
The  DEMNUni simulations, presented in
\cite{Carboneetal15,Castorinaetal15},  are the largest N-body
simulations to date with a massive neutrino component treated as an additional particle type. At present, they are characterised by a
baseline $\Lambda$CDM cosmology to which neutrinos are added with a
degenerate mass spectrum but different total masses. In the near
future, the DEMNUni set will include also an evolving dark energy
background, with different equations of state $w$, in order to study
the degeneracy between $M_\nu$ and $w$, at the non linear level.

The DEMNUni simulations include only Cold Dark Matter (CDM) and
neutrino particles, and have been performed using the tree particle
mesh-smoothed particle hydrodynamics (TreePM-SPH) code
\textsc{gadget-3} \cite{Springel05}, specifically modified in
\cite{Vieletal10} to account for the presence of massive
neutrinos. These simulations  have been run on the Fermi IBM BG/Q
supercomputer at CINECA, Italy\footnote{http://www.hpc.cineca.it/}. 
They contain $(2048)^3$ CDM particles and $(2048)^3$ neutrino particles, in a
comoving cube of volume $V=8\,h^{-3}\textrm{Gpc}^{3}$, and
are characterised by a softening length $\varepsilon=20$ Kpc/$h$. For
each simulation, 62 outputs have been produced, logarithmically equispaced in the scale factor $a=1/(1+z)$, in the redshift interval $z=0-99$, 49 of which lay between $z=0$ and $z=10$.

The DEMNUni set is composed by four cosmological simulations sharing
the same baseline cosmology consistent with the cosmological
parameters estimated by the first Planck data release \citep{Planck13}: $\Omega_{m} = 0.32,\,\Omega_{\Lambda}
=0.68,\,H_{0} =67\,\textrm{km\,s}^{-1}\textrm{Mpc}^{-1},\,\Omega_{b}
=0.05,\,\textrm{and} \, n_{s} = 0.96$.
The only difference between the four simulations is represented by the total neutrino mass $M_\nu$. One of the simulations is characterised by  a pure $\Lambda$CDM cosmology without neutrinos,  and is used as a reference. The other three simulations correspond to a $\Lambda$CDM cosmology plus three degenerate massive neutrinos with $M_\nu=0.17,  0.3, 0.53$ eV. As massive neutrinos modify the
shape of the power spectrum during the cosmic evolution, a special
care has been taken in setting the initial power spectrum in each
simulation. The power spectrum used to set up the  initial conditions
has been obtained from CAMB\footnote{http://camb.info/} adopting the same
primordial scalar amplitude for all the four simulations. On one hand, this ensures that the
normalization of the power spectrum in the four runs is consistent
with Planck, on the other hand it guarantees that the shape of the
initial power spectrum is consistent with the considered neutrino
mass. Given the assumed Planck cosmology, the particle number, and
the volume of the simulations, the CDM mass resolution is about
$8\times10^{10}\,h^{-1}$M$_{\odot}$, adjusted accordingly to the value
of the neutrino particle mass, in order to hold $\Omega_{m} = 0.32$ fixed.
\section{Results}
\label{sec:results}
\subsection{Power spectra and covariance matrices}
\label{sec:covariances}
The power spectra are computed using the standard estimator defined as
\beq
\hat{P}(k) = \frac{1}{VN_{k}}\sum_{k'}|\delta(k')|^{2},
\enq
where $V$ is the survey volume and the sum runs over the $N_{k}$
Fourier modes associated to the $k$-th power spectrum bin.
We measure the real-space power spectra of both the density, $P$, and of the non-linear transform A, $P_{A}$, defined as
\beq
A \equiv \ln(1+\delta), \,\,\, \delta=\frac{\rho - \bar{\rho}}{\bar{\rho}}
\enq
on a $512^{3}$ grid with the nearest-grid
point density assignment deconvolving for the pixel window over the
range $0.0035 \lesssim k \lesssim 0.4$ $h$Mpc$^{-1}$.
The power spectrum covariance matrix is defined as
\beq
\textrm{Cov}_{ij} = \langle \hat{P}(k_i) \hat{P}(k_j) \rangle - \langle  \hat{P}(k_i)
\rangle \langle \hat{P}(k_j) \rangle.
\enq
We follow \cite{Carronetal14d} who developed an approximate form of
the latter in the mildly non-linear regime based on previous 
studies from $N$-body simulations \citep{Neyrinck11, Mohammedetal14}
\beq \label{Cov}
\textrm{Cov}_{ij} =  \delta_{ij} \frac{2 P(k_i)^2}{N_{k_i}} + \sigma_{min}^{2}P(k_i)P(k_j),
\enq
where the first term corresponds to the Gaussian covariance and the second
term approximates the shell-averaged trispectrum of the field.
\cite{Carronetal14d} have shown that the parameter $\sigma_{min}^{2}$ can be interpreted as the
minimum variance achievable on an amplitude-like parameter. It can be
further decomposed into two contributions $\sigma_{min}^{2} =
\sigma_{SS}^{2} + \sigma_{IS}^{2}$,
where the first term is due to the correlation between large
wavelength ``super-survey'' modes with small scales and
the second term corresponds to coupling between small scales or
``intra-survey'' modes.

As we consider the local density fluctuations,
$\delta=\frac{\rho-\bar{\rho}}{\bar{\rho}}$, defined with respect to
the local observed density (i.e the global density fluctuations
divided by the background mode),  we have an additional contribution
to the covariance matrix that can be entirely absorbed within
$\sigma_{SS}^{2}$. The latter can thus be expressed as
$\sigma_{SS}^{2} = (26/21)^{2}\, \sigma_{V}^{2}$,
where following \cite{Takadaetal13}
\beq
\sigma_{V}^{2} = \int P^{L}(k)
W^{2}(\textbf{k}) \frac{d^{3}k}{(2\pi)^{3}} \,.
\enq
In this equation, $W(\textbf{k})$ is the Fourier transform of the
survey volume window function and $P^{L}$ is the linear power spectrum \citep[][for
details]{Carronetal14d, Takadaetal13}. Here $W(x)$ is defined as a spherical
top-hat on the volume $V$.

The contribution to the covariance matrix coming from
$\sigma_{IS}^{2}$ could be more complex, as it was only tested at $z=0$
\citep{Mohammedetal14}.
However, \cite{Carronetal14d} derived an analytical approximation
based on the hierarchical \textit{Ansatz} \citep{Peebles1980, Fry84,
  Bernardeau96} and found that
\beq
\sigma_{IS}^{2} \simeq \frac{P(k_{max})}{V}(4R_{a}+4R_{b}).
\label{eq:sigma2is}
\enq
We assume this expression holds for all redshifts and use it to estimate
the $\sigma_{IS}^{2}$ contribution. Our results are presented in
Table~\ref{table:sigvalues}, and from it $\sigma_{min}^{2}$ is
calculated using the above definition. 
\begin{table}
\centering
\caption{Contributions to the covariance matrix from ``super-survey''
  and ``intra-survey'' modes respectively as a function of redshift. The amplitude of both
  parameters are estimated using the above definitions of
  $\sigma_{SS}^2$ and $\sigma_{IS}^2$.
}
\begin{tabular}{cccc}
\hline
\hline
Redshift &  $\sigma_{SS}^{2}$ & $\sigma_{IS}^{2}$\\
\hline
$z = 0$ & $2.3 \times 10^{-6}$ &  $1.4 \times 10^{-6}$\\
$z = 0.5$ & $1.4 \times 10^{-6}$ &  $6.0 \times 10^{-7}$\\
$z = 1$ & $9.0 \times 10^{-7}$ &  $3.2 \times 10^{-7}$\\
$z = 1.5$ & $5.7 \times 10^{-7}$ &  $2.0 \times 10^{-7}$\\
$z = 2$ & $4.1 \times 10^{-7}$ &  $1.3 \times 10^{-7}$\\
\hline
\hline
\end{tabular}
\label{table:sigvalues}
\end{table}
To quantify its Fisher information content, we need an estimation of
the $A$-covariance matrix as well. We choose to adopt a Gaussian description
for the $A$ field
\beq
\textrm{Cov}^{A}_{ij} = \frac{2}{N_{k}} P_{A}(k_i)
P_{A}(k_j)\delta_{ij}.
\label{eq:covAs}
\enq
\cite{Neyrinck11} have shown this form is valid to a very
good approximation for $k
\lesssim 0.4$ $h$Mpc$^{-1}$, scale at which the covariance
matrix of $P_{A}$ starts to have non-negligible off-diagonal elements. We thus make our
measurements up to this resolution limit, letting us safely assume
that 
$P_{A}$ has a diagonal covariance matrix over the full range of $k$
considered here.
\begin{table}
\centering
\caption{Information gain on neutrino mass for different redshifts. It
  represents the ratio between the elements of the power
  spectrum and $A$-power spectrum Fisher matrix. This gain can be
  interpreted as an effective gain in
the survey volume.}
\begin{tabular}{cccccc}
\hline
\hline
 &  $z=0$& $z=0.5$&$z=1$& $z=1.5$& $z=2$\\
\hline
Gain ($M_\nu =0.235$) & 8 & 5 & 4 & 2.5 & 2.1\\
Gain ($M_\nu =0.415$) & 6.5 & 4 & 3.5 & 2.5 & 2.1\\
\hline
\hline
\end{tabular}
\label{table:gainvalues}
\end{table}
\begin{figure}
\begin{center}
\includegraphics[width=8cm]{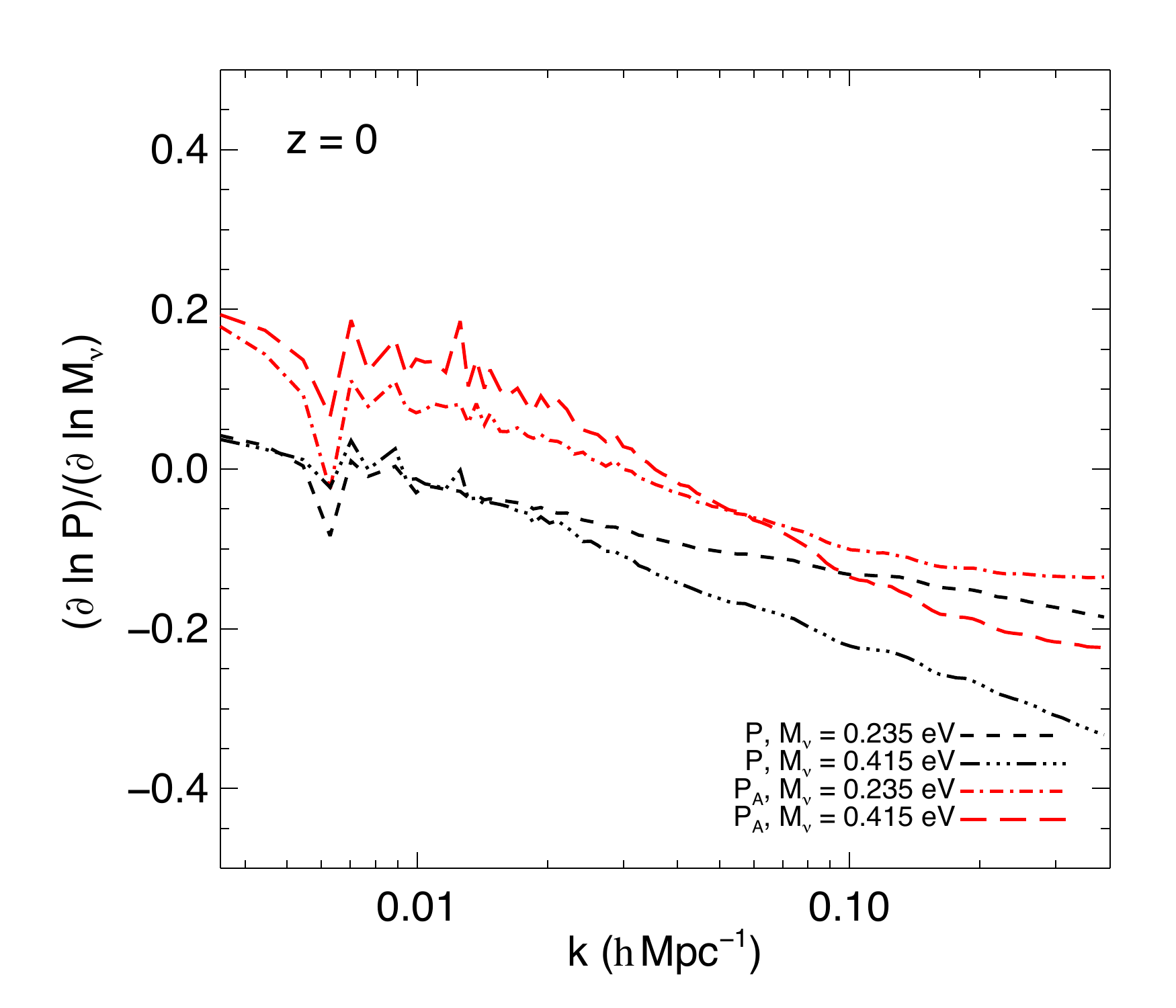}
\end{center}
\caption{Logarithmic derivatives with respect to $M_\nu $ at $z=0$ of
  the power
  spectrum (black lines) and the $A$-power spectrum (red lines) at two
  different fiducial neutrino masses.}
\label{fig:deriv}
\end{figure}
\subsection{Fisher information}
\label{sec:fisherinfo}
Given a set of parameters ${\alpha, \beta, ...}$, the Fisher matrix of
the matter power spectrum is defined as
\beq
F_{\alpha \beta} = \sum_{k_i,k_j < k_{max}} \frac{\partial P(k_i)}{\partial \alpha} Cov^{-1}_{ij}\frac{\partial P(k_j)}{\partial \beta}.
\enq
\cite{Carronetal14d} have shown that the information content of the
power spectrum can be written as
\beq
F_{\alpha \beta} = F^G_{\alpha \beta} - \sigma_{min}^2
\frac{F^G_{\alpha \ln A_{0}}F^G_{\ln A_{0} \beta}}{1 + \sigma_{min}^2
  (S/N)_G^{2}},
\label{eq:infocontent}
\enq
where
\beq
\begin{split}
 F^G_{\alpha \beta} =  \int \frac{\partial  \ln
   P(k)}{\partial \alpha}\frac{\partial  \ln P(k)}{\partial \beta}
 w(k)\, d\ln k \,,
\label{eq:Gterm}
\end{split}
\enq
with $w(k) = V\,k^3/(2\pi)^{2}$.
The discrete sums have been replaced by integrals using the fact that
the number of modes $N_{k}$ is approximately the surface of the shell used for the
bin averaging divided by the distance element between two discrete
modes $N_{k} \simeq V\,(4\pi k^{2}dk)/(2\pi)^{3}$.
Moreover, in Equation~\ref{eq:infocontent}, by analogy to
\cite{Carronetal14d}, we have introduced, for notational convenience, a nonlinear amplitude parameter,
$\ln A_{0}$, defined such as $ \partial_{\ln A_{0}} P(k) = P(k)$. This
parameter corresponds to the initial amplitude $\sigma_{8}^{2}$ in the
linear regime and at $z=0$.
We further define the Gaussian signal to noise as
\beq \label{SNG}
\lp S/N\rp^2_{G} = \int w(k)\, d\ln k \,.
\enq
Using Equation~\ref{eq:covAs}, the Fisher information in the $A$-power
spectrum is given by $F^{A}_{\alpha \beta} = F^{G, A}_{\alpha \beta}$,
where
\beq
 F^{G, A}_{\alpha \beta} =  \int \frac{\partial
\ln P_{A}(k)}{\partial \alpha}\frac{\partial  \ln P_{A}(k)}{\partial
   \beta} w(k)\, d\ln k \,.
\label{eq:infoPa}
\enq
In this analysis we focus on the information about the neutrino mass,
we thus consider $\alpha=\beta=\ln(M_\nu)$ and hold fixed all the other cosmological parameters.

The derivatives in Equations~\ref{eq:Gterm} and \ref{eq:infoPa} are calculated numerically by comparing the power
spectra among simulations with different total neutrino masses. The
derivatives at $z=0$ for both the density and the $A$ fields are shown
in Figure~\ref{fig:deriv} at the minimum and maximum neutrino masses considered:
$M_\nu=(0.17+0.30)/2=0.235$ eV and $M_\nu=(0.30+0.53)/2=0.415$ eV. 

\begin{figure*}
  \begin{center}
    \includegraphics[width=7cm]{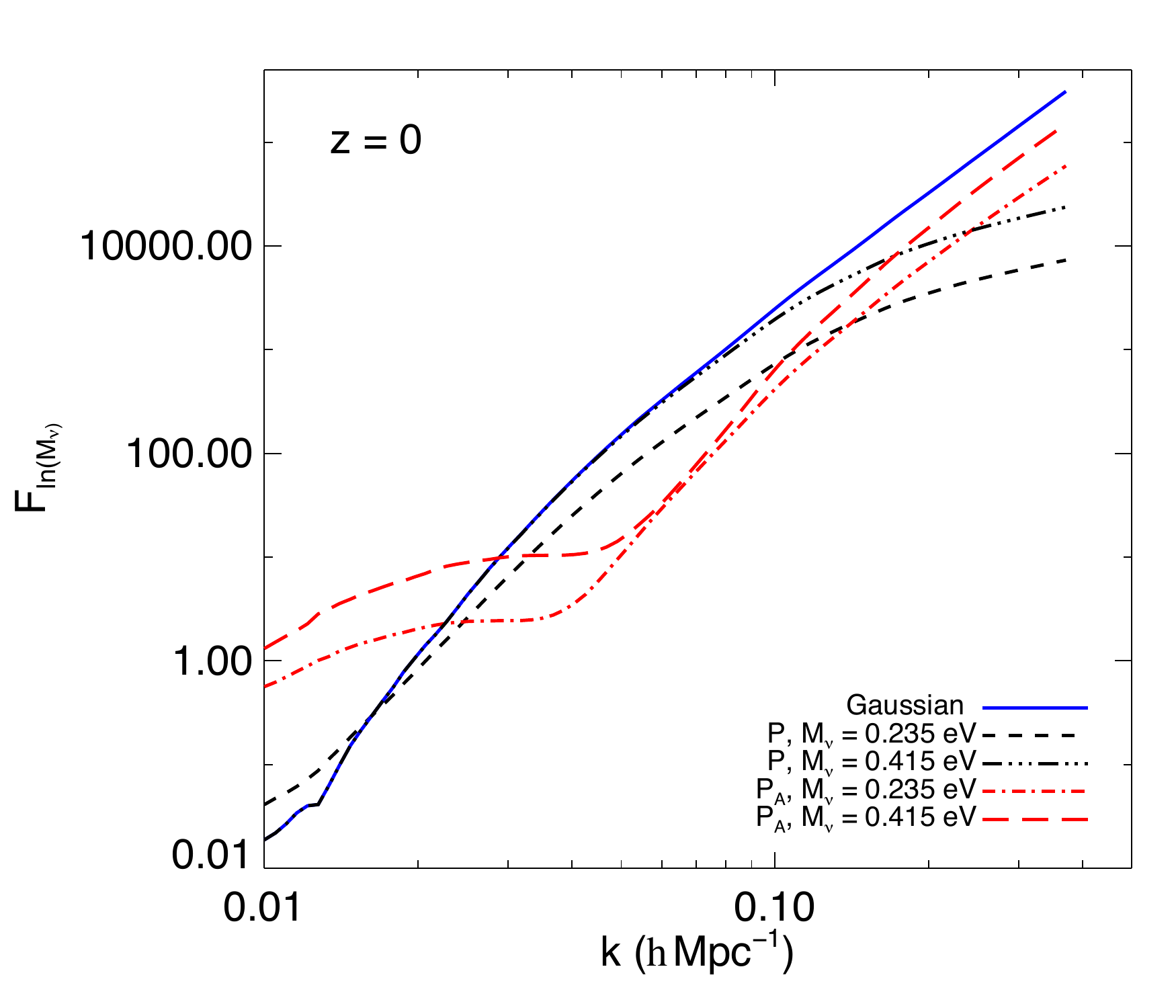}
   \includegraphics[width=7cm]{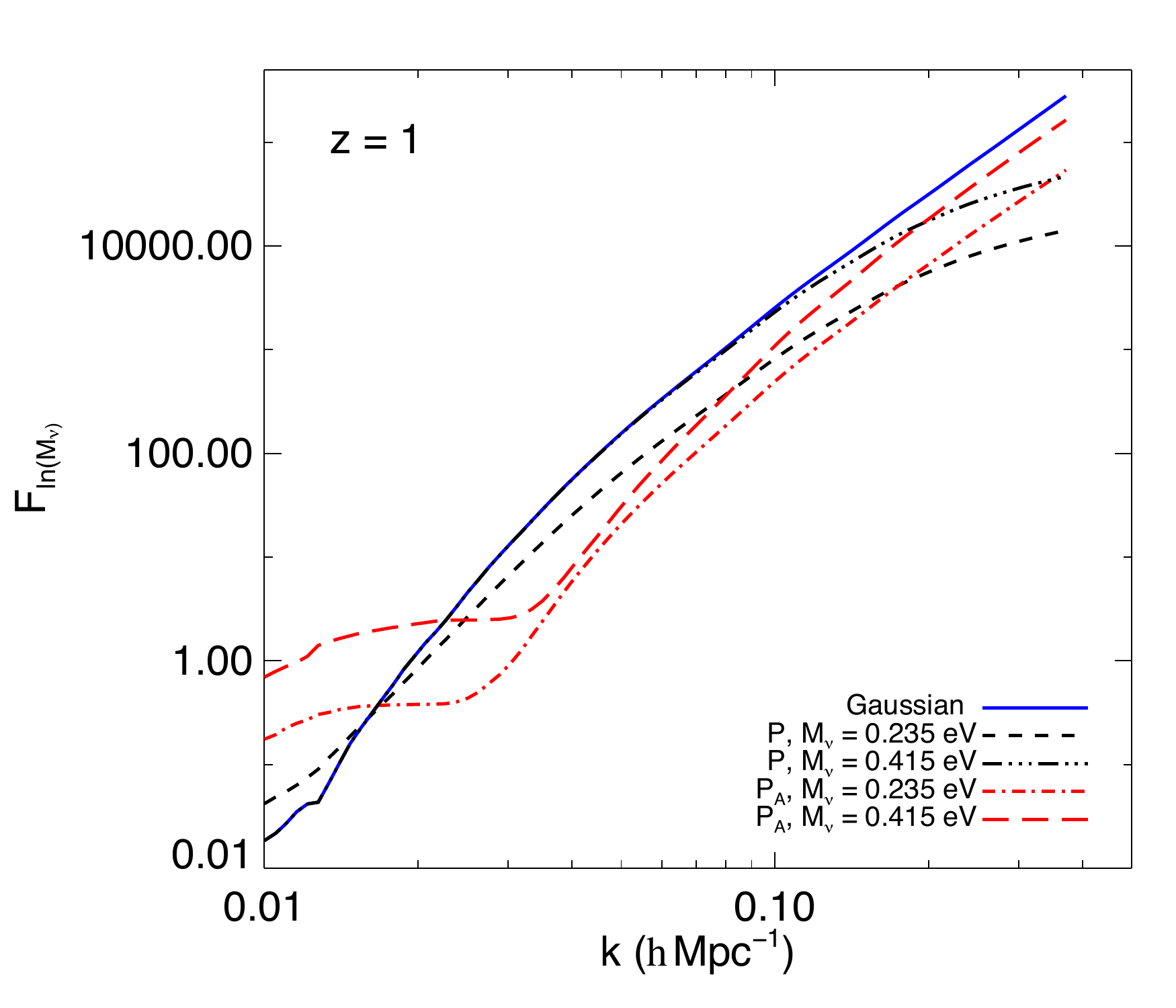}
\end{center}
\caption{Cumulative Fisher information on the logarithm of the
  neutrino mass using both the power
  spectrum (black lines) and the $A$-power spectrum (red lines) for
  two different fiducial values of the neutrino mass. The covariance
  matrices on the two observables are given by
  Equation~\ref{Cov} and Equation~\ref{eq:covAs} respectively. The
  Gaussian part of the matter power spectrum in the case $M_\nu                                                                       
  =0.415$ eV is given for comparison by
the blue solid line. The left panel shows that using the ``sufficient
statistics'' at $z=0$
increases by $\sim 8$ and $\sim 6.5$ the available information for
$M_\nu = 0.235$ eV and $M_\nu = 0.415$ eV
respectively. The right panel shows the same cumulative Fisher
information at $z=1$. At this higher redshift, the gain using
the ``sufficient statistics'' is about $\sim 4$ and $\sim 3.5$ for
$M_\nu = 0.235$ eV and $M_\nu =0.415$ eV. The evolution of
the information gain with redshift is summarized in
Table~\ref{table:gainvalues}.}
\label{fig:Fnu}
\end{figure*}

The results are presented in Figure~\ref{fig:Fnu} which shows the
cumulative Fisher information on the logarithm of the neutrino mass
using both the power spectrum (black lines) and the $A$-power spectrum
(red lines) at two different redshifts: $z=0$ (left panel) and $z=1$ (right panel).
The Gaussian part of the matter power spectrum is given for comparison by
the blue solid line in the case $M_\nu = 0.415$ eV. At our resolution limit, using sufficient statistics
increases by up to $\sim 8$ the available information, leading to error bars
that are almost a factor of $3$ smaller compared to the
matter power spectrum at $z=0$. The evolution of the information gain with the
redshift is presented in Table~\ref{table:gainvalues}. As was shown
previously by \cite{Wolketal14} in the case of projected field, the gain is larger
at small redshift where the non-linearities are stronger. The fiducial
value of the neutrino mass used to estimate the derivatives has almost no
impact on the final gain with, however, a small tendency to have lower
gain at higher neutrino mass as it can be seen on Figure~\ref{fig:Fnu}
or in Table~\ref{table:gainvalues}. Our results show that even at $z=2$, the
sufficient statistics unveils up to $\sim 2$ times more
information compared to the matter power spectrum. These conclusions
directly depend on the value of $\sigma_{min}^{2}$ and a change of
$10\%$ of the latter will result in a change of about
$\sim 5\%$ in the information gain.

It can be seen on Figure~\ref{fig:Fnu} that in the regime $0.025
\lesssim k \lesssim 0.1$ $h$Mpc$^{-1}$, $P_{A}$ is less optimal than
the matter power spectrum. This is due to the bias between $P_{A}$ and $P$ that could
be, in a first approximation, written as $P_{A} = e^{-\sigma_{A}^2}
P_{lin}$ where $\sigma_{A}^2$ is the variance of the $A$-field
\citep{Neyrincketal09}. As this bias depends on the cosmology, it
shifts the $0$-crossing scale of the derivatives of
Equation~\ref{eq:infoPa} from a
scale $k \sim k_{min}$ for the matter power spectrum to $k \sim 0.03$
$h$Mpc$^{-1}$ (see Figure~\ref{fig:deriv}), resulting in a knee in the information. Moving to
larger $k$, the derivatives increase again in absolute value and
$P_{A}$ starts again to perform better than the matter power spectrum.
\section{Discussion}
\label{sec:discuss}
It has been extensively shown that the power spectrum is not the
optimal observable
to constrain cosmological parameters in the mildly non-linear regime as
the information it contains saturates at a finite plateau thus contradicting the
naive Gaussian expectation \citep{Rimesetal05, Rimesetal06,
  Neyrincketal06, Neyrincketal07, dePutteretal12, Takadaetal13}.
Sufficient statistics have been introduced to overcome this issue
and recapture, by design, in their power spectrum most of the
available information.
We have demonstrated, using the DEMNUni simulations, which correspond
to a $\Lambda$CDM cosmology with three degenerate massive neutrinos with
sum of masses respectively 0.17, 0.3 and 0.53 eV, that the power
spectrum of the non-linear transform $A$ outperforms by a factor $\sim
8$ the usual power spectrum, when extracting cosmological information
on the
neutrino mass at $z=0$, and by a factor of $\sim 2$ even at $z=2$. This gain can be seen as an effective gain in
the survey volume. 

Observational issues were not considered in this work, and refinements
need to be made in order to apply the non-linear transform $A$ to actual galaxy survey data.
The first one is the discreteness of galaxy surveys that must be
taken into account.
The effect of the shot-noise on $F_{\alpha, \beta}$ will be to change
$w(k)$ into
\beq
w(k) = V\frac{k^3}{(2\pi)^{2}} \,\,\, \rightarrow w(k) =
V\frac{k^3}{(2\pi)^{2}} \Big( \frac{P(k)}{P(k) +
  \frac{1}{\bar{n}}} \Big)^{2},
\enq
and $P(k) \rightarrow P_{g} = b^{2} P(k)$, where $b$ is the linear galaxy bias.
In the case of the ``sufficient statistics'' a new observable
$A^{\ast}$, optimized for both the non-linearities and the
observational noise, has to be used to recapture the cosmological
information \citep{Carronetal14a, Wolketal14}.
\cite{Wolketal15} have derived in the case of projected fields, the
information content of $A^{\ast}$ as a function of both the information in
the galaxy power spectrum and the cosmological
dependencies of the bias between $P$ and $P_{A}$.
Assuming this approach holds in the 3-dimensional case, we consider
the Sloan Digital Sky Survey (SDSS) LRGs sample number density, $\bar{n} \sim 5.10^{-4}$
$h^{3}$Mpc$^{-3}$ with a galaxy bias of $b = 2$ \citep{Percivaletal07,
  dePutteretal12}. At our $k_{max}$, we find 1$\sigma$ error
bars of $\sim0.01$ and $\sim0.005$ using $P$ and $P_{A^{\ast}}$
respectively leading to an information gain on $M_{\nu}$
of $\sim4$ at $z=0.3$ (corresponding to the peak
of the distribution). As the clustering information content
is well understood in 2D, this simple estimation is expected to
give a realistic order of magnitude, however further investigations
about the relationship between the matter and the non-linear transform
$A$ power spectra need to be made in 3-dimensions.

In this work, we have considered the case of the density fluctuation $\delta$
defined with respect to the local observed density as it is the case for
galaxy surveys. However, in the case of fluctuations defined with
respect to the global density, for example in weak lensing surveys, our
predictions for the information change. The contribution from the intra-survey modes becomes
negligible compared to the one from the super-survey modes, as the
latter can now be written as $\sigma_{SS}^{2} = (68/21)^2\,\sigma_{V}^{2}$.
This gives a value of $\sigma_{SS}^{2}=1.6 \times 10^{-5}$ and $3.0
\times 10^{-6}$ at redshifts $z=0$ and $z=2$, respectively.
 The expected gain in information about the neutrino mass using
$P_{A}$ becomes about factors of $\sim 25$ and $\sim 8$, respectively, compared to the power
spectrum. This behavior was expected as the information in the matter
power spectrum roughly saturates at $1/\sigma_{min}^{2}$. In the
global case, the ``super-survey'' modes dominate resulting in an information
plateau which is lower compared to the local case and thus allowing the non-linear
transform $A$ to perform even better.

The effect of the galaxy bias was already taken into account within the framework of
the ``halo model'' for projected fields \citep{Wolketal14}, as well as
the bias between $P_{A^{\ast}}$ and $P_{g}$. This needs to be extended to the 3-dimensional case.
For future galaxy survey applications, it is also crucial to investigate the effect of redshift
distortions which is left for a future work.

Despite the need of a more sophisticated analysis for a direct application to real
galaxy surveys, this work has demonstrated that
``sufficient statistics'' are expected to be a powerful method to
improve the future constraints on different cosmological
parameters and, in particular, on the neutrino mass.

MW, IS and JC acknowledge NASA grants NNX12AF83G and NNX10AD53G for support.
CC and JB acknowledge financial support from the European Research
Council through the Darklight Advanced Research Grant (n. 291521). 
The DEMNUni simulations were performed on Fermi, the Tier-0  IBM
Blue-Gene/Q system at the ``Centro Interuniversitario del Nord-Est per
il Calcolo Elettronico'' (CINECA, Bologna, Italy),  via the
computational time budget awarded through the class--A call of the
Italian SuperComputing Resource Allocation (ISCRA).
JB, CC and IS would like to thank the DARKLIGHT workshop, ``Measuring and Modelling Redshift-Space Distortions in Galaxy Surveys'', 21-25 July 2014, Sesto (Italy), where part of this work was conceived.
{\small
\bibliographystyle{mn2e}
\bibliography{Nu.bib}

\begin{thebibliography}{}

\bibitem[\protect\citeauthoryear{{Ahmed} et~al.,}{{Ahmed}
  et~al.}{2004}]{Ahmedetal04}
{Ahmed} S.~N.  et~al., 2004, Physical Review Letters, 92, 181301

\bibitem[\protect\citeauthoryear{{Araki} et~al.,}{{Araki}
  et~al.}{2005}]{Arakietal05}
{Araki} T.  et~al., 2005, Physical Review Letters, 94, 081801

\bibitem[\protect\citeauthoryear{{Bernardeau}}{{Bernardeau}}{1996}]{Bernardeau96}
{Bernardeau} F.,  1996, \aap, 312, 11

\bibitem[\protect\citeauthoryear{{Beutler} et~al.,}{{Beutler}
  et~al.}{2014}]{Beutleretal14}
{Beutler} F.  et~al., 2014, \mnras, 444, 3501

\bibitem[\protect\citeauthoryear{{Carbone}, {Petkova} \& {Dolag}}{{Carbone}
  et~al.}{2015}]{Carboneetal15}
{Carbone} C.,  {Petkova} M.,    {Dolag} K.,  2015, in preparation

\bibitem[\protect\citeauthoryear{{Carron} \& {Szapudi}}{{Carron} \&
  {Szapudi}}{2013}]{Carronetal13}
{Carron} J.,  {Szapudi} I.,  2013, \mnras, 434, 2961

\bibitem[\protect\citeauthoryear{{Carron} \& {Szapudi}}{{Carron} \&
  {Szapudi}}{2014}]{Carronetal14a}
{Carron} J.,  {Szapudi} I.,  2014, \mnras, 439, L11

\bibitem[\protect\citeauthoryear{{Carron}, {Wolk} \& {Szapudi}}{{Carron}
  et~al.}{2014}]{Carronetal14d}
{Carron} J.,  {Wolk} M.,    {Szapudi} I.,  2014, ArXiv e-prints

\bibitem[\protect\citeauthoryear{{Castorina}, {Carbone}, {Bel}, {Sefusatti} \&
  {Dolag}}{{Castorina} et~al.}{2015}]{Castorinaetal15}
{Castorina} E.,  {Carbone} C.,  {Bel} J.,  {Sefusatti} E.,    {Dolag} K.,
  2015, in preparation

\bibitem[\protect\citeauthoryear{{de Putter} et~al.,}{{de Putter}
  et~al.}{2012}]{dePutteretal12}
{de Putter} R.  et~al., 2012, \apj, 761, 12

\bibitem[\protect\citeauthoryear{{Dolgov}}{{Dolgov}}{2002}]{Dolgov02}
{Dolgov} A.~D.,  2002, \physrep, 370, 333

\bibitem[\protect\citeauthoryear{{Eguchi} et~al.,}{{Eguchi}
  et~al.}{2003}]{Eguchietal03}
{Eguchi} K.  et~al., 2003, Physical Review Letters, 90, 021802

\bibitem[\protect\citeauthoryear{{Fry}}{{Fry}}{1984}]{Fry84}
{Fry} J.~N.,  1984, \apj, 279, 499

\bibitem[\protect\citeauthoryear{{Hearin}, {Zentner} \& {Ma}}{{Hearin}
  et~al.}{2012}]{Hearinetal12}
{Hearin} A.~P.,  {Zentner} A.~R.,    {Ma} Z.,  2012, \jcap, 4, 34

\bibitem[\protect\citeauthoryear{{Joachimi} \& {Taylor}}{{Joachimi} \&
  {Taylor}}{2011}]{Joachimietal11}
{Joachimi} B.,  {Taylor} A.~N.,  2011, \mnras, 416, 1010

\bibitem[\protect\citeauthoryear{{Lesgourgues} \& {Pastor}}{{Lesgourgues} \&
  {Pastor}}{2006}]{Lesgourguesetal06}
{Lesgourgues} J.,  {Pastor} S.,  2006, \physrep, 429, 307

\bibitem[\protect\citeauthoryear{{Lesgourgues} \& {Pastor}}{{Lesgourgues} \&
  {Pastor}}{2012}]{Lesgourguesetal12}
{Lesgourgues} J.,  {Pastor} S.,  2012, ArXiv e-prints

\bibitem[\protect\citeauthoryear{{McKeown} \& {Vogel}}{{McKeown} \&
  {Vogel}}{2004}]{McKeownetal04}
{McKeown} R.~D.,  {Vogel} P.,  2004, \physrep, 394, 315

\bibitem[\protect\citeauthoryear{{Mohammed} \& {Seljak}}{{Mohammed} \&
  {Seljak}}{2014}]{Mohammedetal14}
{Mohammed} I.,  {Seljak} U.,  2014, \mnras, 445, 3382

\bibitem[\protect\citeauthoryear{{Neyrinck}}{{Neyrinck}}{2011}]{Neyrinck11}
{Neyrinck} M.~C.,  2011, \apj, 736, 8

\bibitem[\protect\citeauthoryear{{Neyrinck} \& {Szapudi}}{{Neyrinck} \&
  {Szapudi}}{2007}]{Neyrincketal07}
{Neyrinck} M.~C.,  {Szapudi} I.,  2007, \mnras, 375, L51

\bibitem[\protect\citeauthoryear{{Neyrinck}, {Szapudi} \& {Rimes}}{{Neyrinck}
  et~al.}{2006}]{Neyrincketal06}
{Neyrinck} M.~C.,  {Szapudi} I.,    {Rimes} C.~D.,  2006, \mnras, 370, L66

\bibitem[\protect\citeauthoryear{{Neyrinck}, {Szapudi} \& {Szalay}}{{Neyrinck}
  et~al.}{2009}]{Neyrincketal09}
{Neyrinck} M.~C.,  {Szapudi} I.,    {Szalay} A.~S.,  2009, \apjl, 698, L90

\bibitem[\protect\citeauthoryear{{Peebles}}{{Peebles}}{1980}]{Peebles1980}
{Peebles} P.~J.~E.,  1980, {The large-scale structure of the universe}

\bibitem[\protect\citeauthoryear{{Percival} et~al.,}{{Percival}
  et~al.}{2007}]{Percivaletal07}
{Percival} W.~J.  et~al., 2007, \apj, 657, 645

\bibitem[\protect\citeauthoryear{{Planck Collaboration} et~al.,}{{Planck
  Collaboration} et~al.}{2014}]{Planck13}
{Planck Collaboration} et~al., 2014, \aap, 571, A16

\bibitem[\protect\citeauthoryear{{Planck Collaboration} et~al.,}{{Planck
  Collaboration} et~al.}{2015}]{Planck15}
{Planck Collaboration} et~al., 2015, ArXiv e-prints

\bibitem[\protect\citeauthoryear{{Rimes} \& {Hamilton}}{{Rimes} \&
  {Hamilton}}{2005}]{Rimesetal05}
{Rimes} C.~D.,  {Hamilton} A.~J.~S.,  2005, \mnras, 360, L82

\bibitem[\protect\citeauthoryear{{Rimes} \& {Hamilton}}{{Rimes} \&
  {Hamilton}}{2006}]{Rimesetal06}
{Rimes} C.~D.,  {Hamilton} A.~J.~S.,  2006, \mnras, 371, 1205

\bibitem[\protect\citeauthoryear{{Seo}, {Sato}, {Dodelson}, {Jain} \&
  {Takada}}{{Seo} et~al.}{2011}]{Seoetal11}
{Seo} H.-J.,  {Sato} M.,  {Dodelson} S.,  {Jain} B.,    {Takada} M.,  2011,
  \apjl, 729, L11

\bibitem[\protect\citeauthoryear{{Springel}}{{Springel}}{2005}]{Springel05}
{Springel} V.,  2005, \mnras, 364, 1105

\bibitem[\protect\citeauthoryear{{Takada} \& {Hu}}{{Takada} \&
  {Hu}}{2013}]{Takadaetal13}
{Takada} M.,  {Hu} W.,  2013, \prd, 87, 123504

\bibitem[\protect\citeauthoryear{{Viel}, {Haehnelt} \& {Springel}}{{Viel}
  et~al.}{2010}]{Vieletal10}
{Viel} M.,  {Haehnelt} M.~G.,    {Springel} V.,  2010, \jcap, 6, 15

\bibitem[\protect\citeauthoryear{{Wolk}, {Carron} \& {Szapudi}}{{Wolk}
  et~al.}{2014}]{Wolketal14}
{Wolk} M.,  {Carron} J.,    {Szapudi} I.,  2014, ArXiv e-prints

\bibitem[\protect\citeauthoryear{{Wolk}, {Carron} \& {Szapudi}}{{Wolk}
  et~al.}{2015}]{Wolketal15}
{Wolk} M.,  {Carron} J.,    {Szapudi} I.,  2015, ArXiv e-prints

\end{thebibliography}
}
\end{document}